\documentclass[11pt,twoside]{article}

\usepackage{asp2004}
\usepackage{epsf}
\usepackage{psfig}
\usepackage{lscape}
\usepackage{graphicx}

\newcommand{\lessim}{$^{<}_{\sim}$}

\begin{document}

\title{The O--C diagram of the subdwarf B pulsating star HS~2201+2610:
detection of a giant planet?}

\author{R. Silvotti$^{1}$,
        S. Schuh$^{2}$,
        R. Janulis$^{3}$,
        S. Bernabei$^{4}$,
        R. \O stensen$^{5}$,\\
	J.-E. Solheim$^{6}$, 
	I. Bruni$^{4}$, 
	R. Gualandi$^{4}$, 
	T. Oswalt$^{7}$, 
	A. Bonanno$^{8}$,\\
	B. Mignemi$^{8}$ and 
	the Whole Earth Telescope Xcov23 collaboration}
	
\affil{$^1$\,INAF-Osservatorio Astronomico di Capodimonte, Napoli, Italy\\
       $^2$\,Institut f\"ur Astrophysik, Universit\"at G\"ottingen, Germany\\
       $^3$\,Institute~of~Theoretical~Physics~\&~Astron.,~Vilnius~University,~Lithuania\\
       $^4$\,INAF-Osservatorio Astronomico di Bologna, Italy\\
       $^5$\,Instituut voor Sterrenkunde, Katholieke Universiteit Leuven, Belgium\\
       $^6$\,Institute of Theoretical Astrophysics, Oslo University, Norway\\
       $^7$\,Florida Institute of Technology, Melbourne, USA\\
       $^8$\,INAF-Osservatorio Astronomico di Catania, Italy}

\begin{abstract}
In this article we present the O--C diagram of the hot subdwarf B pulsating
star HS~2201+2610 after seven years of observations.
A secular increase of the main pulsation period, \.{P}=
(1.3$\pm$0.1)$\times$10$^{-12}$, is inferred from the data.
Moreover, a further sinusoidal pattern suggests the presence of a low-mass
companion (Msin$i$~$\simeq$~3.5 M$_{\it Jup}$), orbiting the hot star at a 
distance of about 1.7~AU with a period near 1140 days.
\end{abstract}

\vspace{-6mm}

\section{Introduction}

The O--C (Observed - Calculated, see e.g. Sterken 2005) diagram can be used to 
measure the secular variation of pulsation periods in time.
This technique is particularly attractive for two classes of pulsators: the hot
subdwarfs (sdBs) and the white dwarfs (WDs) (Kepler et al. 1999, 2005), as
their short pulsation periods (few minutes) are extremely stable in time.
Thus, for these objects, we can obtain very precise O--C measurements that can
open the possibility of detecting secondary bodies with very low masses, 
down to the brown dwarf (BD) and even the planetary limit.
Note that to use more ``traditional'' methods to detect BDs or planets around 
sdBs or WDs is quite difficult. Due to the high gravity, the absorption lines of
these objects are very broad and therefore the radial velocity measurements can
rarely reach the high accuracy required. 
In principle, the transit method can be used, but the probability that a 
transit occurs is very low because of the small radii of these stars 
($\approx$10$^5$ km for an sdB star and $\approx$10$^4$ km for a WD).
Therefore the O--C diagram seems the most promising method to find planets 
around sdB or WD stars.

\begin{table}[]
\begin{center}
\caption[]{Observing log}
\begin{tabular}{lccc} 
\hline
\noalign{\smallskip}
\noalign{\smallskip}
\bf Epoch & \bf telescope & \bf \#\,runs & \multicolumn{1}{c}{\bf \#\,hours}\\
\noalign{\smallskip}
\hline
\noalign{\smallskip}
\noalign{\smallskip}
1999 Oct     & NOT 2.6m$^4$                            & \hspace{1.5mm}1 & \hspace{1.5mm}0.7\\
2000 Sep-Oct & multi-site campaign (7 sites)$^*$       &              24 &              77.3\\
2000 Nov     & CA 2.2m$^4$                             & \hspace{1.5mm}4 & \hspace{1.5mm}8.9\\
2000 Dec     & Loi 1.5m$^3$                            & \hspace{1.5mm}3 & \hspace{1.5mm}6.2\\
2001 May     & Loi 1.5m$^3$, SARA 0.9m$^4$             & \hspace{1.5mm}3 & \hspace{1.5mm}8.0\\
2001 Jun     & Loi 1.5m$^3$                            & \hspace{1.5mm}4 & \hspace{1.5mm}5.3\\
2001 Jul     & CA 1.2m$^4$, NOT 2.6m$^4$               & \hspace{1.5mm}5 &              10.0\\
2001 Aug     & Loi 1.5m$^3$                            & \hspace{1.5mm}9 &  	          48.0\\
2001 Oct-Nov & Loi 1.5m$^3$                            & \hspace{1.5mm}4 & \hspace{1.5mm}8.0\\
2001 Dec     & Loi 1.5m$^3$                            & \hspace{1.5mm}2 & \hspace{1.5mm}3.8\\
2002 May     & Loi 1.5m$^3$                            & \hspace{1.5mm}1 & \hspace{1.5mm}0.6\\
2002 Jul     & La Palma 1.0m$^4$                       & \hspace{1.5mm}3 & \hspace{1.5mm}4.2\\
2002 Aug     & Mol 1.65m$^3$, SLN 0.9m$^1$             & \hspace{1.5mm}5 &              14.0\\
2002 Sep     & Mol 1.65m$^3$                           & \hspace{1.5mm}2 & \hspace{1.5mm}2.8\\
2002 Oct-Nov & Loi 1.5m$^3$, CA 1.2m$^4$, Tenerife 0.8m$^4$ & \hspace{1.5mm}9 &         17.3\\
2003 May-Jun & SARA 0.9m$^4$, Loi 1.5m$^3$             & \hspace{1.5mm}4 & \hspace{1.5mm}4.9\\
2003 Aug-Sep & WET Xcov23 (7 sites)$^{**}$             &              29 &              55.0\\
2003 Sep     & Loi 1.5m$^3$                            & \hspace{1.5mm}2 & \hspace{1.5mm}7.8\\
2004 Jun     & Loi 1.5m$^4$                            & \hspace{1.5mm}7 &              10.8\\
2004 Jul-Aug & Loi 1.5m$^4$, Mol 1.65m$^3$             & \hspace{1.5mm}7 &              24.3\\
2004 Oct     & Loi 1.5m$^4$                            & \hspace{1.5mm}2 & \hspace{1.5mm}2.7\\
2005 Jun     & Loi 1.5m$^4$                            & \hspace{1.5mm}1 & \hspace{1.5mm}1.0\\
2005 Sep  & Mol 1.65m$^3$, Loi 1.5m$^4$, SARA 0.9m$^4$ & \hspace{1.5mm}7 &              38.6\\
2005 Nov-Dec & SARA 0.9m$^4$, Loi 1.5m$^4$             & \hspace{1.5mm}2 & \hspace{1.5mm}3.8\\
2006 Jun     & Loi 1.5m$^4$                            & \hspace{1.5mm}2 &               TBR\\
2006 Jul     & TNG 3.5m$^4$                            & \hspace{1.5mm}5 &               TBR\\
2006 Sep     & CA 2.2m$^4$                             & \hspace{1.5mm}4 &              22.2\\
\noalign{\smallskip}
\hline
\label{obslog}
\end{tabular}
\end{center}
\vspace{-4mm}
{\it Notes:}
CA=Calar Alto, Loi=Loiano, Mol=Moletai, SLN=Serra La Nave.\\
TBR = to be reduced.\\
$^1$~=~1 channel photometer (PMT),
$^2$~=~2 ch. PMT,
$^3$~=~3 ch. PMT,
$^4$~=~CCD.\\
$^*$ {Multi-site campaign: Beijing 0.85m$^3$, Mol 1.65m$^3$, Wendelstein 
0.8m$^4$, Loi 1.5m$^3$, Tenerife 0.8m$^3$, Fick 0.6m$^2$, SARA 0.9m$^4$.}\\
$^{**}$ {Whole Earth Telescope Xcov23: NOT 2.6m$^4$, Lulin 1.0m$^4$, 
OHP 1.9m$^3$, Loi 1.5m$^3$, Wise 1.0m$^4$, Piszkesteto 1.0m$^4$, KPNO 0.4m$^4$.}
\end{table}

\begin{figure}[h]
\centering
\includegraphics[clip,width=12 cm]{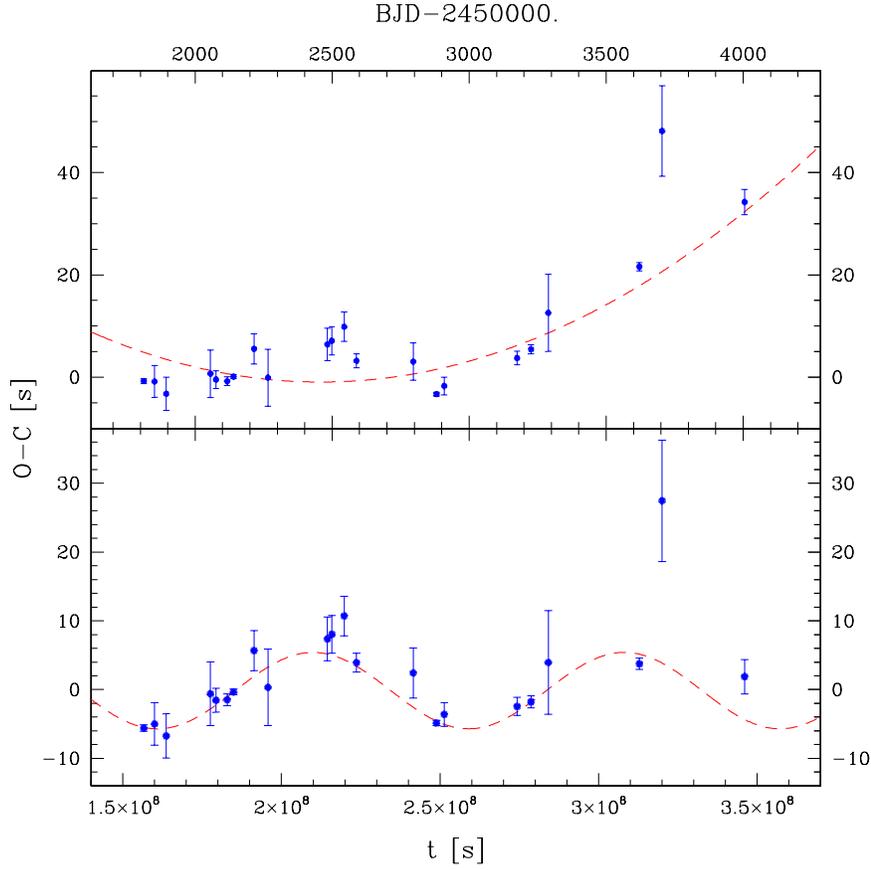}
\caption{The O--C diagram of the main pulsation period of HS~2201+2610.
Upper panel: fit of the long-term trend through a 2nd order polynomial.
Lower panel: residuals and sinusoidal fit.}
\label{O--C}
\end{figure}


\section{The O--C diagram of HS~2201+2610}

The sdB pulsating star HS~2201+2610 (hereafter HS~2201) has been observed for 7
years through a long term monitoring program (151 runs of time-series 
photometry at 17 different telescopes, see Table 1 for more details), with the
main goal of measuring the variation of the pulsation periods in time (dP/dt). 
With a much shorter data set, Silvotti \& Kalytis (2003) were able to 
derive an upper limit $\mid$\.{P}$\mid$$<$3.8$\times$10$^{-12}$ 
relatively to the main pulsation period.
The secular variation of the pulsation periods is related to the structural 
changes in the star due to its evolution. 
When a pulsation period changes in time, the typical shape of the O--C diagram 
is a second-order order polynomial that goes up or down depending whether the 
period is increasing or decreasing.
This effect can be seen in Fig.1 (upper panel), where the O--C diagram suggests 
a value of \.{P}=(1.3$\pm$0.1)$\times$10$^{-12}$.
However, a simple 2nd order polynomial is not able to give a satisfactory fit:
subtracting the polynomial, we still have significant residuals with a 
sinusoidal shape (lower panel of Fig.1) that require a different interpretation.
The simplest explanation is that the sinusoidal behavior is related to a wobble
of the sdB barycenter due to the presence of a low-mass companion.
Depending on its position around the barycenter of the system, the sdB star is
periodically closer or more distant from us by about 5.6 light seconds. 
This is why the timing of the pulsation is cyclicly advanced or delayed.
From our best fit of the O--C diagram and Kepler's third law (assuming a 
circular orbit) we derive: P$_{\it orb}$=1136$\pm$37 days, $a$=$\simeq$1.7~AU,
Msin$i$=$\simeq$3.5 M$_{\it Jup}$ ($a$ being the separation from the star). 
Note that although a projected radius of $\sim$5 light seconds points toward a
planet (for the Sun this value is almost 3~s) however, depending on the 
unknown inclination $i$ of the system, the mass of the companion might also be 
compatible with a BD (2.5$^{\circ}$\lessim$i$\lessim15$^{\circ}$) or even a 
low-mass star ($i$\lessim2.5$^{\circ}$).

\section{Conclusions and next steps}

Although Fig.1 suggests the presence of a low-mass companion around 
HS~2201, the last two points are not consistent with the proposed solution,
suggesting some prudence in our conclusions.
New data are needed in order to confirm this discovery.
Another two runs in June and July 2006 (see Table 1), not yet reduced and 
analyzed, and a new run in November 2006 can help.
On the other hand, the solution proposed in Fig.1 is not the only one and the 
fit can be easily improved if we increase the number of parameters (e.g. 
introducing a second planetary body and/or elliptic orbits).
For all of these reasons the solution proposed in this paper must be considered
only as very preliminary.

Another important point is that HS~2201 has many pulsation periods
(three with amplitudes greater than 1 mma, Silvotti et al. 2002) and it is 
possible, in principle, to build an independent O--C diagram for each of the frequencies.
With new high quality data we aim to obtain the O--C diagram not only of $f$1
(amplitude of 1\%) but also of $f$2 (amplitude of 0.4\%).
Using our present set of data, based on 1-2m class telescopes, the O--C diagram
of $f$2 has very large error bars, but at least its shape is not in 
contradiction with that of $f$1.

On the other hand, to confirm the discovery of a planetary body around
HS~2201 using other methods appears to be rather difficult.
Astrometry and interferometry are excluded due to the distance of this star
(about 1200 pc).
As already discussed in Section 1,
to measure radial velocities with a high accuracy seems difficult due to the 
high surface gravity ($\log{g}$=5.4, \O stensen et al. 2001).
In the IR domain, a typical signature of a cool companion is an IR excess
in the spectral energy distribution. 
This excess is not seen in HS~2201 by the JHK 2MASS measurements, which can 
exclude a late-type MS companion with a mass greater than about 
0.35 M$_{\odot}$.
In order to push down this limit further, we aim to obtain new near-IR 
photometry of higher quality.

\acknowledgements

The authors acknowledge support from
MIUR (COFIN {\it Astrosismologia}, PI L.~Patern\`o);
Deutsche Forschungsgemeinschaft (DFG grant SCHU~2249/3-1);
U.S. National Science Foundation (TDO: AST-0206115).

\end{document}